\documentclass[%
 reprint,superscriptaddress, 
 amsmath,amssymb,
 aps,
]{revtex4-2}
\usepackage{dcolumn}
\usepackage{float}
\usepackage{bm}
\usepackage[T1]{fontenc}
\usepackage{amsmath}
\usepackage{tabu}
\usepackage{booktabs}
\usepackage{tabularx}
\usepackage{multirow}
\usepackage[usenames, dvipsnames]{color}
\usepackage{hyperref}
\usepackage{adjustbox}
\usepackage{tabularx}
\usepackage{graphicx}
\usepackage{bm}
\usepackage{caption}
\begin{document}

\title{Interlayer vibrational hybrid normal mode enabling molecular chiral phonons}

\author{Hanen Hamdi}
\affiliation{Physics Department and IRIS Adlershof, Humboldt-Universit\"at zu Berlin, 12489 Berlin, Germany}
\affiliation{Institute of Physics, Carl von Ossietzky Universit\"at Oldenburg, 26129 Oldenburg, Germany}
\author{Jannis Krumland}
\affiliation{Physics Department and IRIS Adlershof, Humboldt-Universit\"at zu Berlin, 12489 Berlin, Germany}
\author{Ana M. Valencia} 
\affiliation{Institute of Physics, Carl von Ossietzky Universit\"at Oldenburg, 26129 Oldenburg, Germany}
\affiliation{Physics Department and IRIS Adlershof, Humboldt-Universit\"at zu Berlin, 12489 Berlin, Germany}
\author{Carlos-Andres Palma} 
\email{palma@iphy.ac.cn}
\affiliation{Institute of Physics, Chinese Academy of Sciences, P.O. Box 603, Beijing 100190, China}
\author{Caterina Cocchi} 
\email{caterina.cocchi@uni-oldenburg.de}
\affiliation{Institute of Physics, Carl von Ossietzky Universit\"at Oldenburg, 26129 Oldenburg, Germany}
\affiliation{Physics Department and IRIS Adlershof, Humboldt-Universit\"at zu Berlin, 12489 Berlin, Germany}

\date{\today}

\begin{abstract}
Organic/inorganic interfaces formed by monolayer substrates and conjugated molecular adsorbates are attractive material platforms leveraging the modularity of organic compounds together with the long-range phenomena typical of condensed matter. New quantum states are known to be generated by electronic interactions in these systems as well as by their coupling with light. However, little is still known about hybrid vibrational modes.
In this work, we discover from first principles the existence of an infrared-active chiral phonon mode in a pyrene-decorated MoSe$_{2}$ monolayer given by the combination of a frustrated rotation of the molecule around its central axis and an optical mode in the substrate. 
Our results suggest the possibility to enable phonon chirality in molecular superlattices.

\end{abstract}

\maketitle

Hybrid organic/inorganic interfaces formed by low-dimensional transition-metal dichalcogenides (TMDC) and molecular adsorbates are versatile material platforms known to host a variety of new quantum states generated by electronic~\cite{zhen+16nano,amst+19nano,zoje+19ami,krum-cocc21es} and spin interactions~\cite{kafl+17nano,zhao+21jpcl} between the constituents, also triggered by the response to electromagnetic radiation~\cite{liu+17nl,zhan+18am,zhu+18sa}. 
As such, research efforts on these systems may contribute to addressing current challenges in quantum electrodynamics, such as Fano optics~\cite{lien+22nl}, electron-phonon coupling~\cite{reec+20prl}, and molecule–cavity interactions~\cite{jian+22am}.

The recent discovery of chiral phonons in TMDCs~\cite{zhu+18sci,chen+182Dm,jadc+21nano} has opened fascinating scenarios for creating new topological states of matter. 
The first predictions~\cite{zhan-niu15prl,xu+18prb,zhan+18prl} and experiments~\cite{zhu+18sci,jadc+21nano} in this direction have shown that symmetry breaking gives rise to chiral phonon modes with pseudo-spin angular momentum of opposite sign ($l=\pm1$) at opposite valley momenta (\textbf{K} and \textbf{K'}) which can be exploited for valleytronics~\cite{thyg+17,delh+202Dm,zhan+20prb}.
Furthermore, when entangled with single photons, these peculiar quasiparticles are relevant for quantum information technologies~\cite{chen+19natp}.
In this context, molecule-driven vibrational modes in hybrid interfaces would be especially relevant in the design of charge-density-wave phases~\cite{cala-maur11prl,latt+20nano}
and molecular superlattice chiral phonons~\cite{zhan-niu15prl,zhu+18sci}.

In this work, we discover the appearance of an infrared-active rolibrational normal mode in a hybrid interface formed by pyrene molecules deposited on a MoSe$_2$ monolayer. 
This new state of matter appears from the coupling between a frustrated rotation of the molecule and an optical phonon of the TMDC. 
The analysis of the atomic displacements of this vibration in the presence and in the absence of the substrate suggests the intrinsically chiral character of this mode.
We anticipate that chiral phonons ~\cite{zhu+18sci,chen+182Dm,jadc+21nano} are possible in this system. Owing to the non-centrosymmetric adsorption site of pyrene on MoSe$_2$, the clockwise and counterclockwise frustrated rotations of the molecule may define two pairs of chiral superlattice phonons.
These results pave the way for new interlayer topological quantum states.

\begin{figure}
\includegraphics[width=0.45\textwidth]{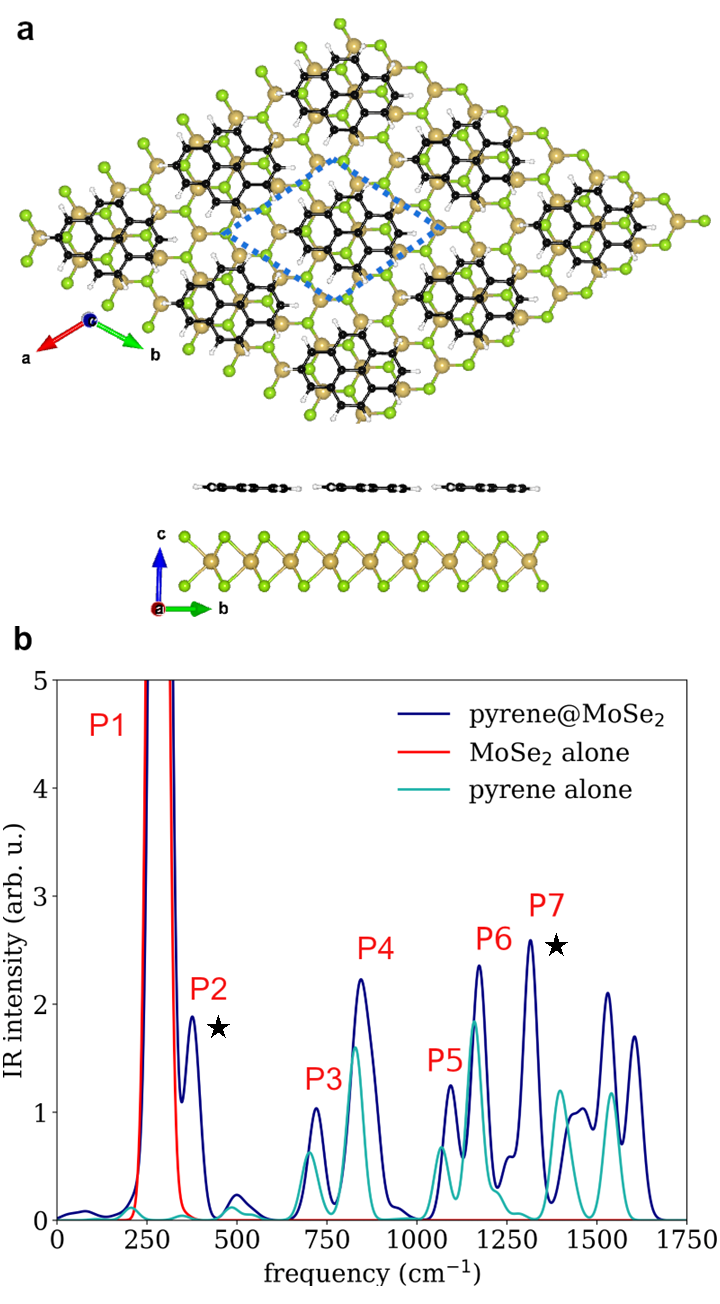}
\caption{ \textbf{a} Ball-and-stick representation (top and side view) of the prototypical hybrid interface formed by pyrene adsorbed on a MoSe$_{2}$ monolayer. Mo, Se, C, and H atoms are depicted in gold, green, black, and white, respectively. In the top panel, the simulation cell is indicated by blue dots. \textbf{b} IR spectra of pyrene@MoSe$_{2}$ and of its isolated constituents. A Gaussian broadening with $\sigma=20$~ cm$^{-1}$ is applied to smear the absorption lines.\label{fig:structure}
}
\end{figure}

As a prototypical system for this investigation, we consider the heterostructure formed by a two-dimensional MoSe$_2$ monolayer in a 3$\times$3 supercell (lattice parameter 3.21~\AA{} in the relaxed geometry) decorated by pyrene molecules adsorbed flat on its basal plane (see Fig.~\ref{fig:structure}\textbf{a}).
In this configuration, the molecules are separated from each other by 2.73~\AA{} (H-H distance); the interlayer separation between the molecules and the TMDC is  3.52~\AA{} (Se-C distance).
All calculations are performed in the framework of density-functional theory~\cite{hohe-kohn64pr,kohn-sham65pr} and density-functional perturbation theory~\cite{gonz95pra,baro+01rmp} (see details in the Supplementary Material).

The vibrational activity of the considered hybrid system is analyzed by means of its infrared (IR) spectrum contrasted against the counterparts obtained for the isolated constituents on the same theoretical footing (see Fig.~\ref{fig:structure}\textbf{b} and Supplementary Material for further details).
These results correspond to phonon modes computed at $\Gamma$. 
For simplicity, the most intense IR modes of the interface are marked by dummy labels: we refer the reader to Table~S3 for the complete list of eigenmodes.

\begin{figure*} 
\centering
  \includegraphics[width=\textwidth]{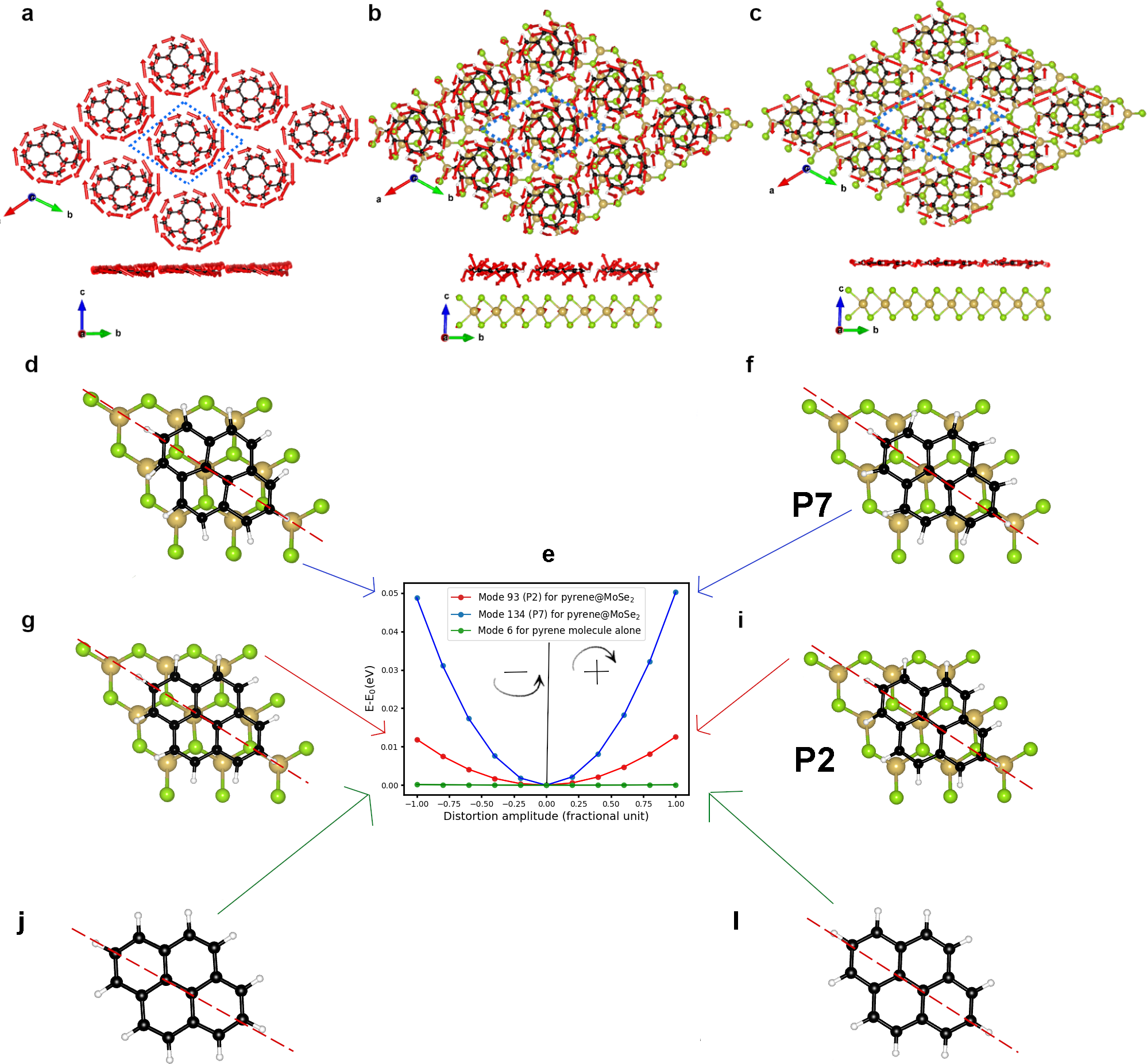}
  \caption{Eigendisplacements (top and side views) of (\textbf{a}) the rotation pyrene in a free-standing  mononolayer and of the modes associated with (\textbf{b})  P2 and (\textbf{c}) P7 in the IR spectrum of the hybrid interface. \textbf{e} PESs computed for the eigenmodes associated with the rotation of a free-standing pyrene monolayer, P2, and P7. Panels \textbf{d} and \textbf{f} depicted the plots of the clockwise ($+$) and anticlockwise ($-$) displacements at distortion amplitudes of 8 fractional units of P7 respectively, panels \textbf{g} and \textbf{i} depicted the plots of the clockwise ($+$) and anticlockwise ($-$) displacements at distortion amplitudes of 8 fractional units of P2 respectively, and panels
\textbf{j} and \textbf{l} show the plots of the clockwise ($+$) and anticlockwise ($-$) displacements 
at distortion amplitudes of 8 fractional units of the  free-standing mononolayer respectivel. The red dashed lines indicate the direction of the long molecular axis after rotation. A large scale of distortion amplitudes of 8 and -8 is shown for better visibility of the displacements.
The reference energy is set at 0 amplitude of the phonon mode, and E$_{0}$ is the ground state energy.
 \label{fig:structure-modes} }
\end{figure*}

The most intense peak visualized in Fig.~\ref{fig:structure}\textbf{b} (P1) is found in the low-frequency region (278.93~cm$^{-1}$) and corresponds to the out-of-plane optical phonon of MoSe$_{2}$ known in the literature as E'~\cite{Mahrouche2022}: it is doubly degenerate and yields the highest IR intensity in the spectrum of MoSe$_{2}$ alone and of its interface with pyrene.  
The fact that P1 appears at the same frequency in the two spectra suggests that this mode is unaffected by the presence of pyrene. 
In the high-frequency part of the IR spectrum of the hybrid interface, the optical modes of pyrene dominate.
Those labeled by P3, P4, P5, and P6 find good correspondence in energy and intensity in the spectrum of pyrene alone (for the analysis of the vibrational modes of pyrene, see Ref.~\cite{herp+21jpca} and references therein).
Frequency shifts are ascribed to expected perturbations due to the presence of the substrate.
The remaining two peaks, P2 and P7, which do not trivially correspond to maxima in the spectra of the single constituents, contain the core results of this work.

P2 at 378.20 cm$^{-1}$ corresponds to a (frustrated) rotation of pyrene coupled with the optical phonon of MoSe$_{2}$. 
It is a new feature of the hybrid system that does not find any correspondence, either at the same frequency or shifted, in the IR spectra of the isolated constituents.
The eigendisplacements of this mode are visualized by red arrows in Fig.~\ref{fig:structure-modes}\textbf{b}. This plot reveals the coupling of the two types of vibrations originating from the organic and inorganic components of the interface and generating the new hybrid mode.
For comparison, we visualize in Fig.~\ref{fig:structure-modes}\textbf{a} the eigendisplacements of the molecular rotation in the free-standing pyrene monolayer. At a glance, the influence of the substrate in the hybrid mode is evident: while the coherent rotational motion of the pyrene molecule is present in both cases (compare Fig.~\ref{fig:structure-modes}\textbf{a} and Fig.~\ref{fig:structure-modes}\textbf{b}), the out-of-plane drifting in the interface is a signature of the coupling with the optical phonon of MoSe$_{2}$. Another IR-active hybrid mode is identified at 1315.81~cm$^{-1}$ (P7). It corresponds to the rotational displacements of the H atoms in pyrene which are activated by the presence of the substrate but that do not couple with it due to the energetic distance from its optical modes (see Fig.~\ref{fig:structure-modes}\textbf{c}).

To deepen our understanding of the interlayer phonon modes, we compute their potential energy surfaces (PES). The results shown in Fig.\ref{fig:structure-modes}\textbf{e} (red and blue curves) indicate that the energy of the system increases with respect to the ground state due to the condensation of stable modes P2 and P7~\cite{Fennie2005, Yajun2018, Mercy2017, Varignon2015, Varignon2019}. 
Moreover, the clockwise (positive) and counterclockwise (negative) displacements are symmetric with respect to the origin.
For comparison, we compute in the same way also the PES associated with the rotational mode of a free-standing pyrene monolayer (Fig.~\ref{fig:structure-modes}\textbf{e}, green curve). The flat profile of the corresponding data points shows the unchanging energy with respect to free rotations confirming that the rotation of the molecules does not correspond \textit{per se} to a vibrational normal mode ~\cite{Cyvin1979,baba2009}.
On the other hand, unlike in free-standing pyrene, the symmetry of the eigendisplacements associated with P2 and P7 (Fig.~\ref{fig:structure-modes}\textbf{f} and Fig.~\ref{fig:structure-modes}\textbf{i}) indicates that the configurations are non-centrosymmetric, such that a clockwise (+) and an anticlockwise (–) chiral displacement can be defined. 

\begin{figure*} 
\centering
  \includegraphics[width=\textwidth]{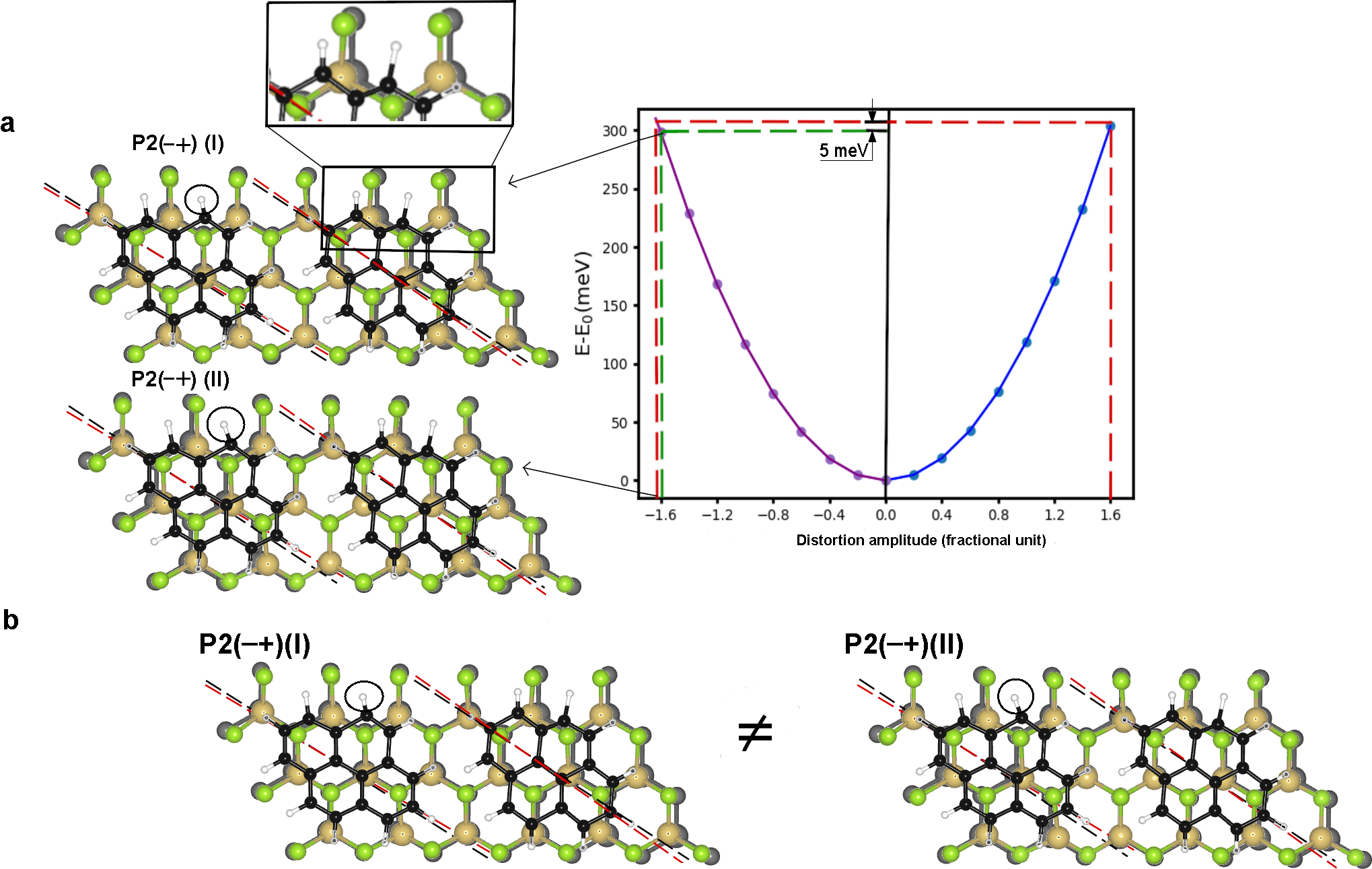}
 \caption{ \textbf{Chiral molecular superlattices}. \textbf{a} The PES calculated for the structure of superlattice reveals that the energies of the displacements for the non-centrosymmetric eigendisplacements of the anticlockwise and clockwise pyrene@MoSe2 displacements ($+-$) and its replica ($-+$) at distortion amplitudes of 1.6 and -1.6 fractional units (fu), respectively are different by 5 meV, since the order of the deformation matters due to the influence of the substrate: the molecules have a larger vibrational amplitude away from each other than towards each other. The left side of the panel \textbf{a} depicted the plots of the P2($-+$) (I) and P2($-+$) (II) structures at -1.6 and -1.62 fu, respectively. \textbf{b} The two dissimilar configurations of P2($-+$) (I) and P2($-+$) (II). The background structure in grey depicts the MoSe$_2$ configuration at the equilibrium position, which is referred to the origin of the atomic positions before the displacements in panel \textbf{a}. The black and red dashed lines indicate the direction of the long molecular axis at the equilibrium position and after rotation, respectively. A large scale of $\pm$8  distortion amplitudes in fu has been applied for the sake of visibility. \label{fig:double-cell} }
\end{figure*}

To simulate a chiral molecular superlattice, we double the extension of the originally considered 3$\times$3 supercell in one in-plane direction (see Fig.~\Ref{fig:double-cell}\textbf{a}). The resulting PES in Fig.~\Ref{fig:double-cell}\textbf{a} shows the case of clockwise–anticlockwise P2($+ -$) vs. anticlockwise–clockwise P2($- +$) displacements, which can be defined for the same eigenmode. 
The chirality of this mode emerges by considering that the same distortion amplitude ($\pm$1.60) leads to a different energy in the PES.
The same energy values are instead obtained for distortion amplitudes of +1.60 and -1.62.
A close inspection of the eigendisplacements associated with the distortion amplitudes of -1.60 and -1.62 points in the PES reveals two different configurations of P2($- +$)(I) and P2($- +$)(II): this can be best seen from the orientation of the H atoms in pyrene with respect to the metal ions in the substrate (see Fig.~\Ref{fig:double-cell}\textbf{d}).

We have checked the robustness of these results with respect to the density of the adsorbed molecules on the MoSe$_2$ monolayer by repeating our initial calculations in a 4$\times$4 supercell (see details in the Supplementary Material).
The hybrid rovibrational normal mode is present also in that heterostructure at a frequency of 278.85~cm$^{-1}$ as found at larger molecular density but its IR intensity is close to zero. 
Moreover, we have checked the role of the specific substrate in the emergence of the hybrid phonon mode by replacing MoSe$_2$ with MoS$_2$. We found that a hybrid normal mode with similar characteristics as those discussed above appears also in the pyrene@MoS$_2$ heterostructure, although it is IR-silent. The results of these additional calculations suggest that rovibrational hybrid modes with chiral character are a common feature of low-dimensional organic/inorganic interfaces.
On the other hand, the IR activity of this mode is very sensitive to the specific characteristics of the hybrid system. 

In summary, using density-functional (perturbation) theory, we have discovered the existence of IR-active rovibrational chiral normal modes in a two-dimensional hybrid heterostructure formed by pyrene molecules adsorbed on a MoSe$_2$ monolayer. 
The analysis of this eigenmode has revealed its hybrid nature emerging from a frustrated rotation of the molecule coupled to an optical phonon mode in the TMDC. 
The chirality of this new phononic state in the hybrid interface has been demonstrated by the analysis of the PES.
Notably, the same rotation in pyrene in the absence of the substrate is non-chiral. 
The presence of this new state of matter has been checked with respect to the chemical nature of the TMDC substrate and to the density of the molecular adsorbates.
The IR activity of this hybrid mode, however, has been found only in the system considered in the main analysis above, leaving the door open for a detailed analysis of the underlying selection rules enabling this characteristic.

The findings reported in this work have relevant implications for the design of quantum materials with tailored topologies hosting new states of matter.
Using molecular adsorbates to obtain chiral phonon modes in TMDCs paves the way for creating superlattices that combine the chemical versatility of carbon-conjugated molecules with the astonishing properties of two-dimensional lattices.
Furthermore, the IR activity of the hybrid mode analyzed in the considered prototypical heterostructure opens fascinating perspectives to trigger and control the rotational motion of adsorbed molecules by means of polarized electromagnetic radiation~\cite{palm+14nl}.

\section*{Acknowledgment}
This work was funded by the German Research Foundation (DFG), project number 182087777 -- CRC 951 and 39064 -- Cluster of Excellence ``Matters of
Activity'', by the German Federal Ministry of Education and Research (Professorinnenprogramm III), by the State of Lower Saxony (Professorinnen für Niedersachsen), and by the Chinese Academy of Sciences (grant numbers QYZDBSSW-SLH038, XDB33000000, XDB33030300). Computational resources were provided by the high-performance computing cluster CARL at the University of Oldenburg, funded by the German Research Foundation (Project No. INST 184/157-1 FUGG) and by the Ministry of Science and Culture of Lower Saxony.


%

\end{document}